# Titanium dioxide (TiO$_2$) Nanoparticles - XRD Analyses – An Insight

**Thirugnanasambandan Theivasanthi*  and Marimuthu Alagar**

Centre for Research and Post Graduate Department of Physics, Ayya Nadar Janaki Ammal College, Sivakasi - 626124, Tamilnadu, India.

**\*Corresponding author:** Phone: +91-9245175532   E-*mail*:  theivasanthi@pacrpoly.org

______________________________________________________________________

**Abstract:** *This work reports aspect related to nano-sized particles of tetragonal anatase phase Titania. This approach is simple, faster, eco-friendly, cost effective and suitable for large scale production. X-Ray Diffraction studies analyze particles size. It is found to be 74 nm and specific surface area is 19.16m$^2$g$^{-1}$. Morphology index (MI) derived from FWHM of XRD data explains the interrelationship of particle size and specific surface area. It is observed that MI has direct relationship with particle size and an inverse relationship with specific surface area. This work throws some light on and helps in the production line of Titania nanoparticles.*

______________________________________________________________________

## 1. Introduction

Titanium dioxide (TiO$_2$) also known as titanium oxide or titanium IV oxide or titania, is the naturally occurring oxide of titanium. It is a versatile transition-metal oxide and a useful material in various present / future applications related to catalysis, electronics, photonics, sensing, medicine, and controlled drug release. Wang *et al*. reported that titania has been extensively studied owing to its physical and chemical properties in photo-catalytic applications for environmental remediation [1]. It is usually used in the form of nanoparticles in suspension for high catalytic surface area and activity [2]. Zallen *et al*. reported that it occurs in nature in anatase, brookite and rutile forms. These phases are characterized with high refractive index (anatase = 2.488, rutile = 2.609, brookite = 2.583), low absorption and low dispersion in visible and near-infrared spectral regions, high chemical and thermal stabilities [3].

In particular, anatase phase is considered for various applications like lithium-ion batteries, filters, anti-reflective & high reflective coatings and has been widely investigated [4]. But, it still remains a challenge to keep this phase stable from easy transformation to rutile. Setiawati *et al*. studied the stabilization of anatase phase with Eu$^{3+}$ and Sm$^{3+}$ ions, added in different concentration [5]. Chen *et al*. report, as a promising photo-catalyst, TiO$_2$ is playing a significant role in helping to solve many serious environmental and pollution challenges. It also bears effective utilization of solar energy based on the photovoltaic and water splitting devices [6].

It is widely used as a photo-catalyst due to its relatively cheap cost, non toxicity and high chemical stability. It has more applications in various industries like aerospace, sports, medicine, paint (to give high gloss, rich depth of color and to replace metal lead), food (to increase the shelf life of products) and cosmetics (UV protection in sunscreens and many other products).

In the proposed study, we have made an attempt to prepare tetragonal pure anatase phase TiO$_2$ nanoparticles in a simple way without any additives and successfully prepared it. Since, no any chemical components used during the preparation of nanoparticles, we believe that the product is biocompatible and bio-safe and can be readily used for food and medicinal industries. Besides, the present method is economical, fast, room temperature, free of pollution, environmentally benign and suitable for large scale production. The particle size of the sample estimated from Debye–Scherer formula (Instrumental broadening) is 74 nm.

## 2. Experimental

Physical grinding method is adopted in this experiment for the size reduction of TiO$_2$ powder. A quantity of 100gm TiO$_2$ powder was put in a grinding machine which having high speed rotator. It was uniformly grinded and crushed well for 15 minutes with utmost precaution to avoid any contamination. At the end, the finely grinded powder was separated. Thus, the bulk TiO$_2$ powder was made as nano-sized powder. The powdered material was packed in plastic pouches and stored in normal room temperature until use.

The XRD analysis of the prepared sample of TiO$_2$ nanoparticles was done using a Bruker make diffractometer, Cu-Kα X-rays of wavelength (λ)=1.5406 Å and data was taken for the 2θ range of 10° to 70° with a step of 0.1972°. The results confirmed the nano sized powder TiO$_2$.

## 3. Results and Discussion

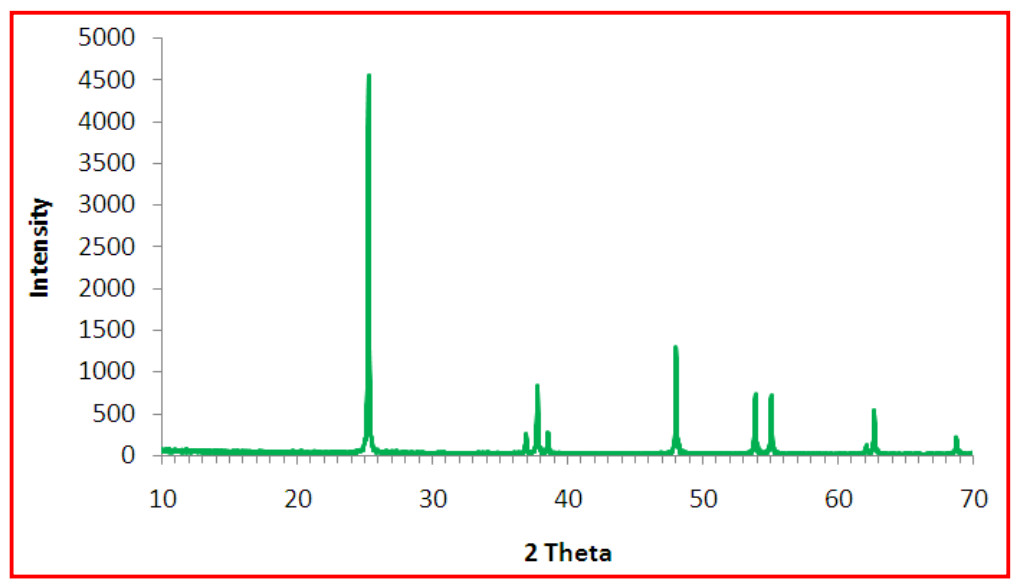

**Fig.1.** XRD pattern of TiO$_2$ nanoparticles

The X-ray diffraction pattern of the synthesized Titania nanoparticles is shown in Fig.1 and the peak details are in Table. 1. Varshney *et al*. reports that absence of spurious diffractions indicates the crystallographic purity [7]. The experimental XRD pattern agrees with the JCPDS card no. 21-1272 (anatase TiO$_2$) and the XRD pattern of TiO$_2$ nanoparticles other literature [8]. The 2θ at peak 25.4° confirms the TiO$_2$ anatase structure [9]. Strong diffraction peaks at 25° and 48° indicating TiO$_2$ in the anatase phase [10]. There is no any spurious diffraction peak found in the sample. The 2θ peaks at 25.27° and 48.01° confirm its anatase structure. The intensity of XRD peaks of the sample reflects that the formed nanoparticles are crystalline and broad diffraction peaks indicate very small size crystallite.

**Table.1.** XRD Data of TiO$_2$ Nanoparticles

| 2θ | θ | Cos θ | Sin θ | FWHM (°) | FWHM Radian | β Cos θ | Size | d-spacing |
|---|---|---|---|---|---|---|---|---|
| 25.270 | 12.635 | 0.97578 | 0.21873 | 0.109 | 0.001902 | 0.001855 | 75 | 3.52149 |
| 36.910 | 18.455 | 0.94857 | 0.31655 | 0.108 | 0.001884 | 0.001787 | 78 | 2.43337 |
| 37.771 | 18.885 | 0.94617 | 0.32366 | 0.122 | 0.002129 | 0.002014 | 69 | 2.37986 |
| 38.528 | 19.264 | 0.94400 | 0.32992 | 0.128 | 0.002234 | 0.002108 | 66 | 2.33482 |
| 48.010 | 24.005 | 0.91350 | 0.40681 | 0.107 | 0.001867 | 0.001705 | 81 | 1.89350 |
| 53.849 | 26.924 | 0.89160 | 0.45280 | 0.122 | 0.002129 | 0.001898 | 73 | 1.70114 |
| 55.037 | 27.518 | 0.88686 | 0.46202 | 0.114 | 0.001989 | 0.001763 | 79 | 1.66720 |
| 62.073 | 31.036 | 0.85684 | 0.51557 | 0.137 | 0.002391 | 0.002048 | 68 | 1.49402 |
| 62.649 | 31.324 | 0.85424 | 0.51987 | 0.130 | 0.002268 | 0.001937 | 72 | 1.48107 |

### 3.1. Particle Size Calculation

From this study, considering the peak at degrees, average particle size has been estimated by using *Debye-Scherer formula*. Inter-planar spacing between atoms (d-spacing) is calculated using *Bragg's Law* and enumerated in Table.1.

$$D = \frac{0.9\lambda}{\beta cos\theta} \quad \text{............(1)}$$

$$2dsin\theta = n\lambda \quad \text{............(2)}$$

Where, λ is wave length of X-Ray (0.1540 nm), β is FWHM (full width at half maximum), θ is diffraction angle, d is d-spacing and D is particle diameter size.

### 3.2. Instrumental Broadening

When particle size is less than 100 nm, appreciable broadening in x-ray diffraction lines will occur. Diffraction pattern will show broadening because of particle size and strain. The observed line broadening will be used to estimate the average size of the particles. The total broadening of the diffraction peak is due to sample and the instrument. The sample broadening is described by

$$FW(S) \times cos\theta = \frac{K \times \lambda}{Size} + 4 \times Strain \times sin\theta \quad \text{............(3)}$$

The total broadening $\beta_t$ equation is described by

$$\beta_t^2 \approx \{\frac{0.9\lambda}{D\cos\theta}\}^2 + \{4\varepsilon\tan\theta\}^2 + \beta_o^2 \quad \ldots\ldots\ldots\ldots\ldots\ldots\ldots\ldots\ldots\ldots(4)$$

Where D is average particle size, $\varepsilon$ is strain and $\beta_0$ is instrumental broadening. The size and strain of the experimentally observed broadening of several peaks are computed simultaneously using *least squares method* and presented in Fig.2. When, particle size becomes smaller, due to size effect, the peaks become broad and widths larger. The broadening of peak may also occur due to micro strains of the crystal structure arising from defects like dislocation and twinning [11].

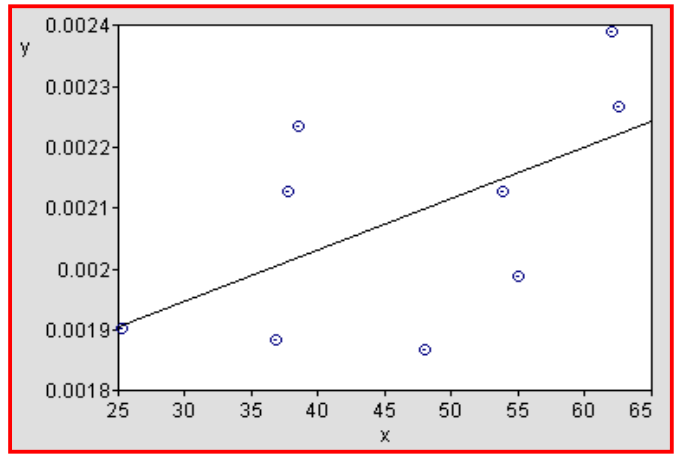

**Fig.2.** Typical Instrumental Broadening

Williamson and Hall proposed a method for deconvoluting size and strain broadening by looking at the peak width as a function of 2θ. W-H plot is shown in Fig.3. It is plotted with sin θ on the x-axis and β cos θ on the y-axis (in radians). A linear fit is got for the data and from it; particle size (74 nm) and strain (0.0000175) are extracted from y-intercept and slope respectively.

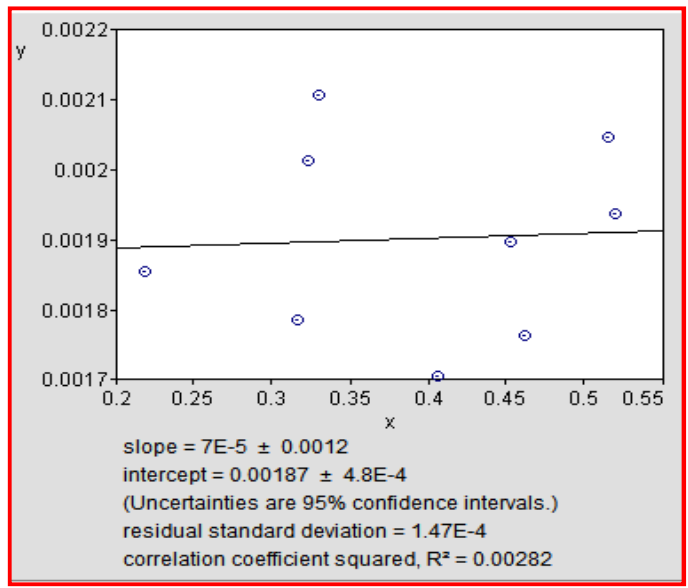

**Fig.3.** Williamson Hall Plot - $TiO_2$ nanoparticles

### 3.3. Specific Surface Area (SSA)

The surface states will play an important role in the nanoparticles, due to their large surface to volume ratio with a decrease in particle size [12]. SSA is a material property. It is a derived scientific value that can be used to determine the type and properties of a material. It has a particular importance in case of adsorption, heterogeneous catalysis and reactions on surfaces. SSA is the Surface Area (SA) per mass.

Zhang *et al*. report, the specific surface area and surface to volume ratio increase dramatically as the size of materials decreases. The high surface area of $TiO_2$ nanoparticles facilitates the reaction / interaction between $TiO_2$ based devices and the interacting media, which mainly occurs on the surface or at the interface and strongly depends on the surface area of the material [13]. Mathematically, SSA can be calculated using formulas (5 & 6) and both formulas yield same result [14-15]. The observed result is in Table 2.

$$SSA = \frac{SA_{part}}{V_{part} * density} \quad \text{............(5)}$$

$$S = 6 * 10^3 / D_p \rho \quad \text{............(6)}$$

Where SSA & S are the specific surface area, $V_{part}$ is particle volume and $SA_{part}$ is Surface Area, $D_p$ is the size (spherical shaped), and $\rho$ is the density of $TiO_2$ (4.23 g.cm$^{-3}$).

**Table.2.** Specific Surface Area of $TiO_2$ Nanoparticles

| Particle Size (nm) | Surface Area (nm$^2$) | Volume (nm$^3$) | Density g.cm$^{-3}$ | SSA (m$^2$ g$^{-1}$) | SA to Volume Ratio |
|---|---|---|---|---|---|
| 74 | 17203.36 | 212175 | 4.23 | 19.16 | 0.08 |

### 3.4. Dislocation Density

The dislocation density is the length of dislocation lines per unit volume of the crystal [16]. A dislocation is a crystallographic defect, or irregularity, within a crystal structure. The presence of dislocation strongly influences many of the properties of materials. Mathematically, it is a type of topological defect. It increases with plastic deformation; a mechanism for the creation of dislocations must be activated in the material. Dislocation formation are formed by three mechanisms i.e. homogeneous nucleation, grain boundary initiation, and interface the lattice and the surface, precipitates, dispersed phases, or reinforcing fibres.

The movement of a dislocation is impeded by other dislocations present in the sample. Thus, a larger *dislocation density* implies a larger hardness. Chen and Hendrickson measured and determined dislocation density and hardness of several silver crystals. They found that crystals with larger dislocation density were harder [17]. It has been shown that the dislocation density increases while the grain size decreases with increasing strain and ultimately these parameters reach saturation values [18].

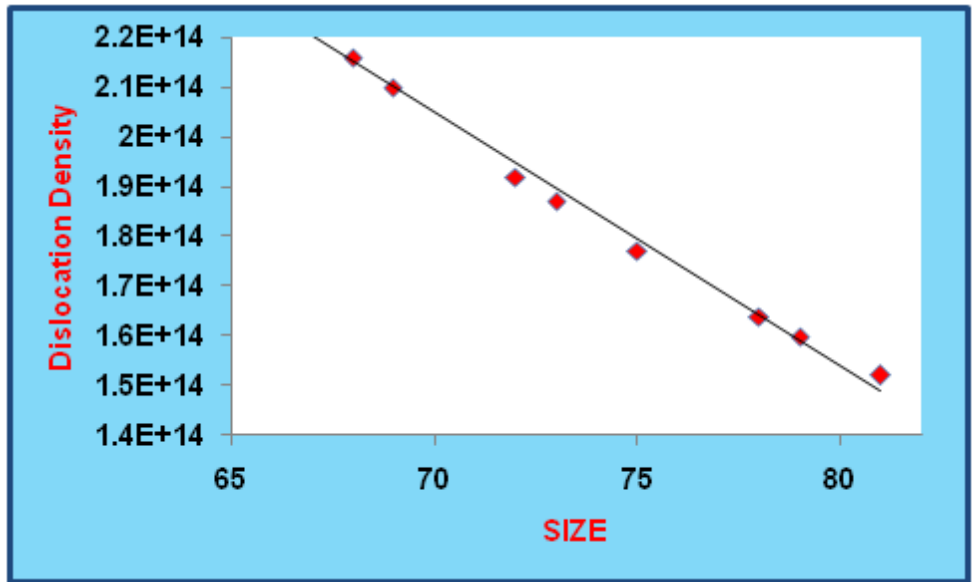

**Fig.4.** Particle Size Vs Dislocation Density of TiO$_2$ Nanoparticles

It is well known that above a certain grain size limit (~20 nm) the strength of materials increases with decreasing grain size [19-20]. The X-ray line profile analysis has been used to determine the intrinsic stress and dislocation density [21-22]. The dislocation density ($\delta$) in the sample has been determined using expressions (7 & 8) and results from both the formulas are approximately same [23-24]. The sample $\delta$ is 1.82 x 10$^{14}$.m$^2$ and the results are enumerated in Table.3.

$$\delta = \frac{15\beta\cos\theta}{4aD} \quad \text{...............................................................................(7)}$$

$$\delta = 1/D^2 \quad \text{......................................................................................(8)}$$

Where, $\delta$-dislocation density, $\beta$- diffraction broadening - measured at half of its maximum intensity (radian), $\theta$- diffraction angle (degree), $\alpha$-lattice constant (nm) and D-particle size (nm). It is observed from the tabulated details and from Fig.4; dislocation density of the sample is indirectly proportional to particle size. Dislocation density increases while particle size decreases. It implies that the prepared TiO$_2$ nanoparticles have more strength and hardness than their bulk (TiO$_2$) counterpart.

### 3.5. Morphology Index (MI)

It is well known that TiO$_2$ nanopowder is widely used in many diverse industries. Such uses are derived from its unique structural, physical and chemical properties, which are reflected by its hardness, surface properties, particle size and morphology. It is proposed that the specific surface area of TiO$_2$ nanopowder depends on the interrelationships of particle morphology and size. MI is calculated from FWHM of XRD to explore this relationship, based on our earlier report [25].

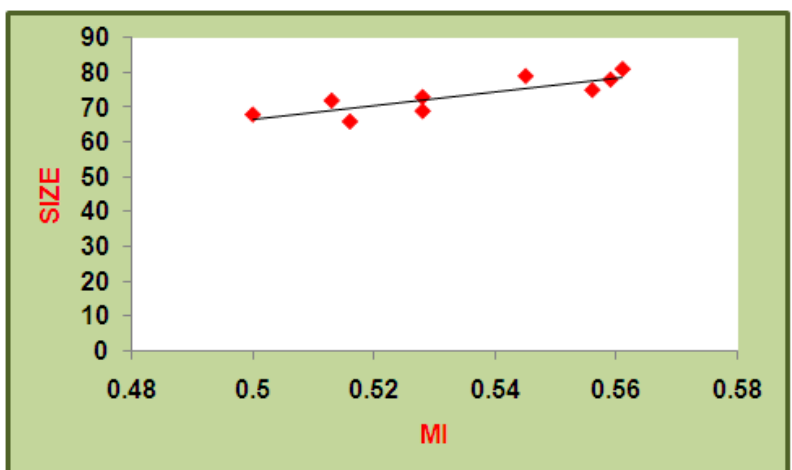

**Fig.5.** Morphological Index Vs Particle Size of TiO$_2$ Nanoparticles

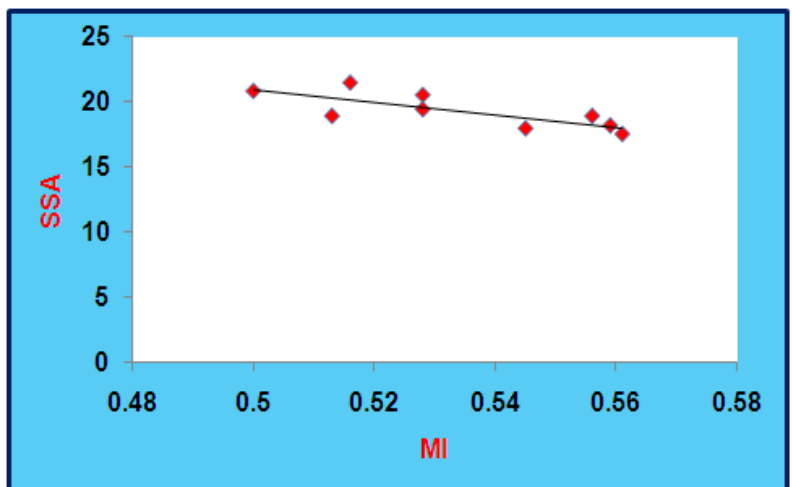

**Fig.6.** MI Vs Specific Surface Area of TiO$_2$ Nanoparticles

**Table.3.** Morphology Index and Dislocation Density of TiO$_2$ Nanoparticles

| 2θ (deg) | FWHM (β) radians | Particle Size (D) nm | Surface Area (nm$^2$) | Volume (nm$^3$) | Specific Surface Area (m$^2$g$^{-1}$) | Morphology Index (unit less) | Dislocation Density (m$^2$) |
|---|---|---|---|---|---|---|---|
| 25.270 | 0.001902 | 75 | 17671.5 | 220893 | 18.91 | 0.556 | 1.77 x 10$^{14}$ |
| 36.910 | 0.001884 | 78 | 19113.4 | 248475 | 18.18 | 0.559 | 1.64 x 10$^{14}$ |
| 37.771 | 0.002129 | 69 | 14957.1 | 172007 | 20.54 | 0.528 | 2.10 x 10$^{14}$ |
| 38.528 | 0.002234 | 66 | 13684.8 | 150533 | 21.49 | 0.516 | 2.29 x 10$^{14}$ |
| 48.010 | 0.001867 | 81 | 20612 | 278262 | 17.51 | 0.561 | 1.52 x 10$^{14}$ |
| 53.849 | 0.002129 | 73 | 16741.5 | 203689 | 19.43 | 0.528 | 1.87 x 10$^{14}$ |
| 55.037 | 0.001989 | 79 | 19606.7 | 258155 | 17.95 | 0.545 | 1.60 x 10$^{14}$ |
| 62.073 | 0.002391 | 68 | 14526.7 | 164636 | 20.85 | 0.500 | 2.16 x 10$^{14}$ |
| 62.649 | 0.002268 | 72 | 16286 | 195432 | 18.91 | 0.513 | 1.92 x 10$^{14}$ |

MI is obtained using equation,

$$M.I = \frac{FWHM_h}{FWHM_h + FWHM_p} \quad \ldots\ldots\ldots\ldots\ldots\ldots\ldots\ldots\ldots\ldots\ldots\ldots(9)$$

Where, M.I. is morphology index, $FWHM_h$ is highest FWHM value obtained from peaks and $FWHM_p$ is particular peak's FWHM for which M.I. is to be calculated.

MI range of experimental $TiO_2$ nanopowder is from 0.50 to 0.56 and the details are presented in Table.3. It is correlated with the particle size (range from 66 to 81 nm) and specific surface area (range from 17.5 - 21.49 $m^2g^{-1}$). It is observed that MI is directly proportional to particle size and inversely proportional to specific surface area with a small deviation. The results are shown in Fig.5 & Fig.6. Linear fit in the figures indicates the deviations and relationships between them. The observed results of the MI confirm the uniformity and fineness of the prepared nanoparticles.

## 4. Conclusions

We have successfully synthesized $TiO_2$ nanoparticles, in a versatile, non-toxic and bio-safe approach, at room temperature without using any chemical components. This method is single source, catalyst free, simple, economic and environmentally benign which will make it suitable for various applications. XRD analyses have confirmed that the synthesized particles are tetragonal anatase phase $TiO_2$ and their nanosize. XRD has also analyzed their various characters like specific surface area, dislocation density and morphology index. It is well expected that this synthesis technique would be extended to prepare many other important metal oxide nanostructures.

**Acknowledgements**

The authors express their immense thanks to VIT University, Vellore for providing XRD instrument to analyze the sample. They also acknowledge assistances and encouragements of staff & management of PACR Polytechnic College, Rajapalayam and Ayya Nadar Janaki Ammal College, Sivakasi, India.